\documentstyle[sprocl]{article}

\newcommand{\ice}[1]{\relax}

\newcommand{\g}{\gamma}

\def\beq{\begin{equation}}
\def\eeq{\end{equation}}
\def\bea{\begin{eqnarray}}
\def\eea{\end{eqnarray}}

\begin{document}
\begin{flushright}
\begin{tabular}{l}
  TTP96--22\\
  MPI/PhT/96-66\\
  hep-ph/9607422\\
  July 1996
\end{tabular}
\end{flushright}
\vskip0.5cm

\title{Higher Moments of Heavy Quark Vacuum Polarization}

\author{ K.G.~Chetyrkin}
\address{  Max-Planck-Institut f\"ur Physik, Werner-Heisenberg-Institut,\\
   F\"ohringer Ring 6, D-80805 Munich, Germany\\
   and\\
   Institute for Nuclear Research, Russian Academy of Sciences,\\
   60th October Anniversary Prospect 7a, Moscow 117312, Russia}
\author{J.H.~K\"uhn,  M.~Steinhauser}
\address{  Institut f\"ur Theoretische Teilchenphysik,\\ 
   Universit\"at Karlsruhe, 
   D-76128 Karlsruhe, Germany}

\maketitle\abstracts{
We present   analytical calculation of the first seven 
moments of the  heavy quark vacuum polarization function  in the  
three-loop order. 
The  obtained results are compared against the 
asymptotic formulas   following  from the threshold singularities.
We also discuss the  $\mu$ dependence of the moments within 
the  BLM procedure. 
                    }
\section{Introduction}\label{sec:intr}
The quark vacuum polarization function
\beq
\Pi^f(q^2) = \frac{-i}{3q^2}\int {\rm d} x e^{iqx}
\langle 0| \;
T\;j_{\mu}^{f}(x)j_{\mu}^{ f }(0)\;|0 \rangle
\label{Pi}
{},
\eeq
with $j^f_\mu = \overline{\psi}_f \g_{\mu}\psi_f $ being the vector current
for a quark $f$, is an interesting object related to a number of
important physical quantities.  An (incomplete) list includes:
\begin{itemize}
\item  
The combination\footnote{We ignore the so-called  singlet contribution 
proportional to $(\sum_f Q_f)^2$ and first appearing  in order $\alpha_s^3$.}
\beq
D(q) = \frac{1}{q^2}\frac{1}{1+e^2\Pi^{em}(q^2)}
\ 
\mbox{\rm  with } 
\ 
\Pi^{em}  = \sum_f Q_f^2 \Pi^f
\label{photon}
\eeq 
is the quark contribution (to order $e^2$) to the photon 
propagator $D(q^2)$.
\item
The optical theorem
relates  the inclusive cross-section of $e^ + e^-$
annihilation into hadrons and
thus the function $R(s) = \sigma_{tot}/\sigma_{point}$
to the discontinuity of $\Pi^{em}$
in the complex plane
\begin{equation} 
\label{disc}
R(s) =  \displaystyle
  12\pi \,{\rm Im}\, \Pi^{em}( s +i\epsilon)
{}\, .
\end{equation}
Conversely,  the vacuum polarization
is obtained through a
dispersion relation from
its absorptive part, vis., 
\begin{eqnarray}
\Pi^{em}(q^2) &=& \frac{q^2}{12\pi^2} \int_{4m^2}^\infty \,ds\,\, 
                \frac{R(s)}{s(s-q^2)} 
{}.
\label{eqdisprel}
\end{eqnarray}
\item
Replacing the electromagnetic 
charge $Q_f$ in $\Pi^{em}$ by the vector weak coupling
$
v_{{f}}=2I_3^{{f}}-4Q_{{f}}\sin^2
\theta_{{\rm w}}
$
one arrives at the vector part of the quark contribution 
to the Z boson propagator and, again through the dispersion relation, 
to the vector part of the decay rate of the $Z$ boson to hadrons. 
\end{itemize}

To order $\alpha_s$ the calculation of $\Pi^f$ was performed by
K\"all\'en and Sabry in the context of QED a long time  ago
\cite{KaeSab55}.  With measurements of ever increasing precision,
predictions in order $\alpha_s^2$ are needed for a reliable comparison
between theory and experiment.

Such a calculation was recently presented in
\cite{CheKueSte95Pade}. It  employed  a  semianalytic  approach   based on 
conformal mapping and Pad\'e approximation
\cite{FleTar94,BroFleTar93,BroBaiIly94,BaiBro95}.  
Important ingredients were the leading terms of the expansion of
$\Pi^f$ in the high energy region $(-q^2)/m^2\gg 1$
\cite{{CheKatTka79DinSap79CelGon80,GorKatLar86}}, the Taylor series
around $q^2=0$ which has been evaluated up to terms of order $(q^2)^4$
and information about the structure of $\Pi$ in the threshold region.
The method was tested in the case of the ``double bubble'' diagrams
against the known analytical result \cite{HoaKueTeu95} with highly
satisfactory results. 
The calculation  leads to a point-wise
prediction of the absorptive part $R^f(s)$ in order $\alpha_s^2$ with
conservatively estimated accuracy of about 6\%,
which should be sufficient 
for comparison with experimental results in the foreseeable future.

An important case when the full mass dependence of $ \Pi^f$ is
necessary is a precise determination of $\alpha_s$ and $m_b$ from QCD
sum rules for $\overline{b} b$ \cite{Voloshin95}. This is because of
the still persisting contradiction between the values of $\alpha_s$
determined from low- and high-energy measurements \cite{bb-crisis}. 
Fortunately enough, QCD
sum rules for $\overline{b} b$ are based on the moments of production
cross section of $\overline{b} b$ states in $e^+ e^-$ annihilation
which can been computed exactly even  in order $\alpha_s^2$.

In the present paper we report on  analytical calculation of the
first seven moments of the function $ \Pi^f$ or, equivalently, of its
Taylor series around $q^2=0$ up to (and including) terms of order
$(q^2)^7$.  

\ice{
It is organized as follows: The notation is presented in
section \ref{sec:not}. The results of the calculation of the moments
are given in section \ref{sec:cal}. Next section deals with the
threshold singularities and asymptotic behaviour of moments at large
$n$.  Choice of the renormalization parameter $\mu$ is discussed in
section \ref{sec:mu}.  Section \ref{sec:concl} contains a brief summary
and conclusions.
}

\section{Notations}\label{sec:not}

We deal with the case of QCD containing $n_l = N_f -1$ massless quarks
and a  massive quark $\psi_F$ with the (pole) mass $ m_F$ later
referred to  as $m$. The  corresponding polarization function
$\Pi^F$ and the ratio $R^F$ will be denoted as $\Pi$ and the ratio $R$,  
respectively. 
It is convenient to define
\begin{eqnarray}
\Pi(q^2) &=& \Pi^{(0)} + \frac{\alpha_s(\mu^2)}{\pi} C_F \Pi^{(1)} 
\\
&&
       + \left(\frac{\alpha_s(\mu^2)}{\pi}\right)^2
         \left[
                 \overbrace{
                C_F^2       \Pi_A^{(2)}
              + C_A C_F     \Pi_{\it NA}^{(2)}
              + C_F T   n_l \Pi_l^{(2)}
              + C_F T       \Pi_F^{(2)}
                }^{\displaystyle\Pi^{(2)}}
         \right].
\nonumber
\end{eqnarray}
with  $\alpha_s(\mu)$ being  the $\overline{\mbox{MS}}$ coupling constant.
For every polarization function $\Pi^{(i)}_?$ with $i=0,1,2$ and
$? =A,NA,l,F$ (if $i=2$ ) we define its moments $C^{(i)}_?$
as follows 
\beq
\Pi^{(i)}_?(q^2) = \frac{3}{16\pi^2}\sum_{n > 0}  C^{(i)}_{?,n} 
\left(\frac{q^2}{4m^2}\right)^n
\label{moms1}
\eeq
or, equivalently, 
\beq
C^{(i)}_{?,n} = \frac{4}{9}\int_{4m^2}^\infty
R^{(i)}_{?,n}(s)\left(\frac{4m^2}{s}\right)^n\frac{d s}{s}
=
\frac{4}{9}
\int_0^1 R^{(i)}_{?,n}(v)\left(1-v^2\right)^{n-1}d (v^2)
{},
\label{moms2}
\eeq
with $v=\sqrt{1-\frac{4m^2}{s}}$. 
Note that by definition  
\beq
C^{(2)}_{n}= C_F^2 C^{(2)}_{A,n} 
+ C_A C_F C^{(2)}_{NA,n}+ C_F T n_lC^{(2)}_{l,n} + C_F T C^{(2)}_{F,n}
{}.
\nonumber 
\eeq
In order to fix the absolute normalization we put below the lowest order
result  for  $R(s)$:
\beq
R^{(0)}(s) = 3 \left(1 + \frac{2 m^2}{s}\right)\sqrt{1-\frac{4 m^2}{s}}
{}.
\eeq
\section{The calculation}\label{sec:cal}
Important information is contained in the Taylor series of
$\Pi(q^2)$ around zero.
The first seven coefficients of the Taylor series around $q^2=0$
are calculated with the help of the program MATAD which has been used 
previously for the three-loop corrections to the $\rho$ parameter
\cite{CheKueSte95rho},
to $\Delta r$
\cite{CheKueSte95deltar}
and the top loop induced reaction $e^+e^- \to H l^+ l^-$
\cite{KniSte95}.
The package is written in FORM 
\cite{VerFORM}.
For a given diagram  it  calculates  the derivatives with respect to
the external momentum $q^2$ to the desired order, raising internal
propagators to powers up to about 20. It then performs the
traces and uses recurrence relations 
based on the integration by parts method
\cite{CheTka81,Bro92}
to reduce the resulting three-loop tadpole diagrams 
(in dimensional regularisation).
In order to save space we present our results converted in numerical form 
in  the Table 1. The original analytical expressions can be found in
our paper   \cite{rqcdlong}.

The coefficients $C_1^{(2)} - C_4^{(2)}$ were already presented in
\cite{CheKueSte95Pade}, $C_5^{(2)}, C_6^{(2)}$ and $C_7^{(2)}$ are new.
The $C_F^2$ term of $C_1^{(2)} - C_3^{(2)}$ and $C_5^{(2)}$ are 
in agreement with~\cite{BaiBro95} and~\cite{Bai96}, respectively.
The results in the Table are given for a particular choice of 
$\mu = m$; in order to transform them to a general $\mu$ the following
relation should be used:
\beq
C^{(2)}_{n}(\mu') = 
C^{(2)}_{n}(\mu) + C^{(1)}_{n}\beta_0 \ln \left(\frac{\mu'}{\mu}\right)^2 
\  \
\mbox{with} 
\  \ 
\beta_0 = \frac{11}{12} C_A - \frac{T}{3}N_f
\label{renormgroup}
\eeq 
\begin{table}[ht]
\renewcommand{\arraystretch}{1.3}
\begin{center}
\begin{tabular}{|l||r|r||r|r|r|r||r|r|}
\hline
  n 
  & $C_n^{(0)}$ 
  & $C_n^{(1)}$ 
  & $C_{A,n}^{(2)}$ 
  & $C_{{\it NA},n}^{(2)}$
  & $C_{l,n}^{(2)}$ 
  & $C_{F,n}^{(2)}$ 
  & $C_n^{(2),5}$ 
  & $C_n^{(2),4}$
 \\
\hline
\hline
1 & 1.067  &   4.049&  5.075& 7.098  &-2.339&0.7270  &31.66&33.22\\
2 & 0.4571 &   2.661&  6.393& 6.311  &-2.174&0.2671 &30.99&32.44\\
3 & 0.2709 &   2.015&  6.689& 5.398  &-1.896&0.1499 &28.53&29.79\\
4 & 0.1847 &   1.63 &  6.685& 4.699  &-1.671&0.0995 &26.29&27.40 \\
5 & 0.1364 &   1.372&  6.574& 4.165  &-1.494&0.0723 &24.41&25.41\\
6 & 0.1061 &   1.186&  6.426& 3.746  &-1.353&0.0557 &22.84&23.74\\
7 & 0.0856 &   1.046&  6.267& 3.409  &-1.239&0.0446 &21.5 &22.33\\
\hline
\end{tabular}
\end{center}
\caption{\label{allmoments} 
The results for the first seven moments.  The last two columns display
the moments $C_n^{(2)}$ for the cases of $N_f=5$ and  $N_f=4$,
respectively.
         }
\end{table}

\section{Higher $n$ moments and threshold singularities}\label{sec:thresh}
In this section we compare our exact moments with
their asymptotic behaviour  originating   from the corresponding threshold 
singularities, with $r^{(i)}_{?,n}$ standing for the ratio
$C^{(i)}_{?,n}/C^{(i),as}_{?,n}$.  

As it follows from (\ref{moms2}) the asymptotic behaviour
of the moments in the limit of large $n$ is completely determined by 
the behaviour of $R(s)$ at the threshold $s \approx 4m^2$.  
The constraints on the threshold behaviour  of $\Pi(q^2)$ originate from 
our knowledge about the nonrelativistic Greens function in the presence
of a Coulomb potential and its interplay with ``hard'' vertex
corrections. For a theory with nonvanishing $\beta$ function
(QED with light fermions or QCD) the proper definition of the
coupling constant and its running  must
be taken into account. 
Below we cite results that are available in the literature about the threshold behaviour 
of $R(s)$ and  its moments. 

\begin{eqnarray}
R^{(0),thr} &=& 3\left(\frac{3 v}{2} -\frac{v^3}{3} \right), 
\\
C^{(0),as}_n&=& \frac{\sqrt{\pi}}{n^{3/2}} -\frac{7\sqrt{\pi}}{8n^{5/2}} 
{}.
\label{thrR0}
\end{eqnarray}

\begin{eqnarray}
R^{(1),thr} &=& 3\left(\frac{3}{4}\pi^2 - 6v \right), 
\\
C^{(1),as}_n &=& \frac{\pi^2}{n} -\frac{4\sqrt{\pi}}{n^{3/2}} 
{}.
\label{thrR1}
\end{eqnarray}

\noindent{\large\bf    $C_F^2$ part} \cite{BarGatKoeKun75,VolSmi94,BaiBro95}:  
\begin{eqnarray}
R_A^{(2),thr} &=& 3\left(\frac{\pi^4}{8v} - 3\pi^2 \right), 
\\
C^{(2),as}_{A,n} &=& \frac{\pi^{9/2}}{6n^{1/2}} - 4\frac{\pi^2}{n}
{}.
\label{thrA}
\end{eqnarray}

\noindent{\large\bf    $C_F C_A$ part} \cite{CheKueSte95Pade}:  
\begin{eqnarray}
R_{\it NA}^{(2),thr}&=&3\pi^2
                         \left(
                              -\frac{11}{16}\ln\frac{v^2 s}{\mu^2}
                              +\frac{31}{48}
                         \right),
\label{thrNA}
\\  
C^{(2),as}_{NA,n} &=& \frac{\pi^2}{12 n}\left(
\frac{31}{3} - 11 \ln  4 + 11 \sum_{i=1}^{n-1} \frac{1}{i}
                           \right)
{}.
\label{CthrNA}
\end{eqnarray}

\noindent{\large\bf    $C_F n_l  T$ part} \cite{HoaKueTeu95}:
\begin{eqnarray}
R_{\it l}^{(2),thr}&=&3\pi^2
                         \left(
                              \frac{1}{4}\ln\frac{v^2 s}{\mu^2}
                              -\frac{5}{12}
                         \right),
\label{thrl}
\\
C^{(2),as}_{l,n} &=& \frac{\pi^2}{3 n}\left(
-\frac{5}{3} +  \ln  4  - \sum_{i=1}^{n-1} \frac{1}{i}
                                      \right)
{}.
\label{Cthrl}
\end{eqnarray}

\noindent{\large\bf    $C_F T$ part} \cite{HoaKueTeu95}:  

\begin{eqnarray}
R_F^{(2),thr} &=&  (22 - 2\pi^2) v  + (-\frac{245}{18} + \frac{4}{3}\pi^2) v^3
\label{thrF}               
{},
\\
C^{(2),as}_{F,n} &=& \frac{\sqrt{\pi}}{9 n^{3/2}}
\left( 
-4 \pi^2 +44 - \frac{172}{3n} + \frac{11\pi^2}{2n}
\right)
{}.
\label{CthrF}
\end{eqnarray}

In  Table 2 we compare the asymptotic formulas with the available moments.
We observe that at $n=7$ the first two terms of the asymptotic expansion  in
$1/n$  agree with the exact results with the accuracy of about 35\%. 

\begin{table}[ht]
\renewcommand{\arraystretch}{1.3}
\begin{center}
\begin{tabular}{|l||r|r||r|r|r||r|r|r|}
\hline
  n 
  & $r_n^{(0)}$ 
  & $r_n^{(1)}$ 
  & $r_{A,n}^{(2)}$ 
  & $r_{{\it NA},n}^{(2)}$
  & $r_{l,n}^{(2)}$ 
  & $r_{F,n}^{(2)}$ 
  & $r_n^{(2),5}$ 
  & $r_n^{(2),4}$
 \\
\hline
\hline
1 & 4.814&   1.457&  -0.474 &    -1.755 &2.536 &2.51 &-0.845&-0.902\\
2 & 1.297&   1.096&  10.51  &    2.522  &1.032 &1.28 &5.522 &4.624  \\
3 & 1.121&   1.046&  1.937  &    1.700    &0.9709&1.128&2.078 &1.982  \\
4 & 1.067&   1.031&  1.479  &    1.499  &0.9611&1.075&1.643 &1.597  \\
5 & 1.043&   1.024&  1.322  &    1.407  &0.9608&1.049&1.472 &1.442  \\
6 & 1.03 &   1.020 & 1.243  &    1.353  &0.9628&1.035&1.38  &1.357  \\
7 & 1.022&   1.018&  1.197  &    1.317  &0.9653&1.026&1.321 &1.304  \\
\hline
\end{tabular}
\end{center}
\caption{\label{moments_over_asympt} 
The results for the first seven ratios of exact moments to their asymptotic values.
The last two column display
the ratios for the case of $N_f=5$ and  $N_f=4$,
respectively.
         }
\end{table}

\section{Choice of $\mu$} \label{sec:mu}

The $\mu$ dependence of $C^{(2)}$ as displayed by eq. (\ref{renormgroup})
is  to the running of $\alpha_s(\mu^2)$.  
Now we want to apply the BLM procedure
\cite{BroLepMac83}
to our results. It suggests to choose the scale of the
${\cal O}(\alpha_s)$ term in such a way that the contribution of the 
${\cal O}(\alpha_s^2)$ part proportional to $\beta_0$ is absorbed.
This prescription is based on the observation that the remaining coefficients
of the $\alpha_s^2$ terms are often relatively small.
It is possible to treat each term of ${\cal O}(z^n)$
separately. In Table \ref{tabblm} we list the BLM scale
$\mu_{BLM}$ for each Taylor coefficient together with the
numerical value of the original ($C_n^{(2)}$) and the
${\cal O}(\alpha_s^2)$ term which remains after the BLM scale 
is adjusted ($C_n^{BLM}$).
\begin{table}[ht]
\renewcommand{\arraystretch}{1.3}
\begin{center}
\begin{tabular}{|l||r|r||r|r|r||r|r|r||r|r|}
\hline
n & $C_n^{(0)}$ & $C_n^{(1)}$ 
  & $C_n^{(2)}$ 
  & $C_n^{BLM}$ & $\mu_{BLM}/m$ 
  & $\bar{C}_n^{(1)}$ & $\bar{C}_n^{(2)}$ \\
\hline
\hline
1 & 1.07 &  4.05 & 30.1 & 13.7 & 0.420 &1.92     & 3.82 \\
2 & 0.46 &  2.66 & 29.5 & 14.3 & 0.294 &0.83    & 3.69 \\
3 & 0.27 &  2.01 & 27.3 & 14.0 & 0.244 &0.39    & 2.50 \\
4 & 0.18 &  1.63 & 25.2 & 13.5 & 0.215 &0.15     & 1.65 \\
5 & 0.14 &  1.37 & 23.4 & 13.0 & 0.195 &0.01    & 1.12 \\
6 & 0.11 &  1.19 & 21.9 & 12.5 & 0.181 &-0.09   & 0.85 \\
7 & 0.09 &  1.05 & 20.7 & 12.0 & 0.169 &-0.15   & 0.76 \\
\hline
\end{tabular}
\end{center}
\caption{\label{tabblm} The BLM scale $\mu_{BLM}$ is expressed in terms of the 
                        original scale $m$. For the numerical values of 
                        $C_n^{(2)}$ and $\bar{C}_n^{(2)}$ 
                        $n_l=5$ and $\mu=m$, respectively, $\mu=\bar{m}$  
                        is used. The double bubble
                        diagrams with two massive fermion loops are also
                        included. $C_n^{BLM}$ is the coefficient remaining
                        after the $\beta_0$ term is absorbed.}
\end{table}
It is interesting to note that $\mu_{BLM}$
is decreasing with increasing $n$. 
This is plausible because the higher Taylor coefficients
are increasingly dominated by
the threshold region and the characteristic scale at
threshold is the relative three-momentum of the quarks.
Note that $C_n^{BLM}$ remains nearly constant whereas $C_n^{(2)}$ 
decreases for increasing $n$ (but remember the rapidly decreasing 
coefficients $C_n^{(0)}$). 

In comparison to $C_n^{(2)}$ in Table \ref{tabblm} also the
corresponding one- and two-loop coefficients ($C_n^{(0)}, C_n^{(1)}$)
and the two- and three-loop coefficients ($\bar{C}_n^{(1)}, \bar{C}_n^{(2)}$)
in the $\overline{\mbox{MS}}$ scheme are listed. Whereas
$C_n^{(1)}$ is roughly of ${\cal O}(1)$ 
$\bar{C}_n^{(1)}$ varies from approximately $1.9$ down to $0.01$.
Also the sign changes. Therefore the BLM procedure is rather unstable and 
not applicable. On the other hand, the 
$\overline{\mbox{MS}}$ coefficients $\bar{C}_n^{(2)}$ are 
already reasonably small, so there is no urgent need for an optimization
procedure. 
\section{Conclusions and summary} \label{sec:concl}
We have  presented analytical  results for first seven 
moments of the  heavy quark vacuum polarization function 
to order ${\cal O}(\alpha_s^2)$.
We found that BLM procedure does not lead to 
a significant improvement of the perturbation theory for the moments.
It is demonstrated that the threshold singularities do describe
the higher moments (with $n=6,7$) with inaccuracy not exceeding 
40\% (in order $\alpha_s^2$).

\section*{Acknowledgments}
\noindent
We are grateful to D.~Broadhurst for helpful discussions.
Comments by S.~Brodsky and M.~Voloshin on the threshold behaviour
are gratefully acknowledged.
One of the authors (K.~Ch.) would like to thank the Organizing
Committee of the second workshop on CONTINUOS ADVANCES IN QCD for the
occasion to present the results of this paper.

The work was supported by BMFT under Contract 056KA93P6, 
DFG under Contract Ku502/6-1, INTAS under Contract INTAS-93-0744
and by the Graduiertenkolleg Elementarteilchenphysik,
Karlsruhe.
\section*{References}
\sloppy
\raggedright
\def\app#1#2#3{{\it Act. Phys. Pol. }{\bf B #1} (#2) #3}
\def\apa#1#2#3{{\it Act. Phys. Austr.}{\bf #1} (#2) #3}
\def\lhc{Proc. LHC Workshop, CERN 90-10}
\def\npb#1#2#3{{\it Nucl. Phys. }{\bf B #1} (#2) #3}
\def\plb#1#2#3{{\it Phys. Lett. }{\bf B #1} (#2) #3}
\def\prd#1#2#3{{\it Phys. Rev. }{\bf D #1} (#2) #3}
\def\pR#1#2#3{{\it Phys. Rev. }{\bf #1} (#2) #3}
\def\prl#1#2#3{{\it Phys. Rev. Lett. }{\bf #1} (#2) #3}
\def\prc#1#2#3{{\it Phys. Reports }{\bf #1} (#2) #3}
\def\cpc#1#2#3{{\it Comp. Phys. Commun. }{\bf #1} (#2) #3}
\def\nim#1#2#3{{\it Nucl. Inst. Meth. }{\bf #1} (#2) #3}
\def\pr#1#2#3{{\it Phys. Reports }{\bf #1} (#2) #3}
\def\sovnp#1#2#3{{\it Sov. J. Nucl. Phys. }{\bf #1} (#2) #3}
\def\jl#1#2#3{{\it JETP Lett. }{\bf #1} (#2) #3}
\def\jet#1#2#3{{\it JETP Lett. }{\bf #1} (#2) #3}
\def\zpc#1#2#3{{\it Z. Phys. }{\bf C #1} (#2) #3}
\def\ptp#1#2#3{{\it Prog.~Theor.~Phys.~}{\bf #1} (#2) #3}
\def\nca#1#2#3{{\it Nouvo~Cim.~}{\bf #1A} (#2) #3}
\def\ijmp#1#2#3{{\it Int.~J~.~Mod.~Phys.~}{\bf #1} (#2) #3}
\def\mpl#1#2#3{{\it  Mod.~Phys.~Lett.~}{\bf #1} (#2) #3}

\end{document}